# The log-periodic-AR(1)-GARCH(1,1) model for financial crashes

L. Gazola [a], C. Fernandes [a], A. Pizzinga [b] and R. Riera [c, 1]

[a]Departamento de Engenharia Elétrica, Pontifícia Universidade Católica do Rio de Janeiro, 22453-900 Rio de Janeiro, RJ ,Brasil

[b] Instituto de Gestão de Riscos Financeiros e Atuariais (IAPUC), Pontifícia Universidade Católica do Rio de Janeiro, 22453-900 Rio de janeiro, RJ, Brasil

[c]Departamento de Física, Pontifícia Universidade Católica do Rio de Janeiro, 22453-900 Rio de Janeiro, RJ, Brasil

## Abstract

This paper intends to meet recent claims for the attainment of more rigorous statistical methodology within the econophysics literature. To this end, we consider an econometric approach to investigate the outcomes of the log-periodic model of price movements, which has been largely used to forecast financial crashes. In order to accomplish reliable statistical inference for unknown parameters, we incorporate an autoregressive dynamic and a conditional heteroskedasticity structure in the error term of the original model, yielding the log-periodic-AR(1)-GARCH(1,1) model. Both the original and the extended models are fitted to financial indices of U. S. market, namely S&P500 and NASDAQ. Our analysis reveal two main points: (i) the log-periodic-AR(1)-GARCH(1,1) model has residuals with better statistical properties and (ii) the estimation of the parameter concerning the time of the financial crash has been improved.

**Short title**: The log-periodic-AR(1)-GARCH(1,1) model

---

[1] Corresponding author: rrif@fis.puc-rio.br

# 1. Introduction

Some years ago, Sornette et al. [1] suggested that, prior to crashes, the mean function of the index price time series is characterized by a power law acceleration decorated with log-periodic oscillations, leading to a finite-time singularity that describes the onset of the market crash. Within this model, this behavior would hold for months and years in advance, allowing the anticipation of the crash from the log-periodic oscillations exhibited by the prices. The heuristic hypothesis of the model is the existence of a growing cooperative trading action of the market agents due to an imitative behavior among them. In the pre-crash regime, clusters of correlated trades with arbitrary sizes would drive the financial system, and therefore, the observed financial variables would exhibit scaling invariant properties. The emergence of a financial crash would then be analogous to a critical point of a standard second-order phase transition of physical systems, with the log-periodic oscillations being signatures of the discrete scaling symmetry of the underlying informational network of the market.

Since then, several authors have reported a large number of empirical results for a variety of unrelated crashes in worldwide stock markets indices. For a recent review of the theoretical framework of the log-periodic model and a compilation of empirical evidences, see Ref. [2]. Although there have been some efforts to perform statistical tests of detection of log-periodicity [3-6] so far in the literature, many investigations carried out to establish evidence for log-periodic oscillations in price time series were based on direct curve fitting and call for a proper statistical inference analysis.

This paper intends to meet recent concerns about the need of more rigorous and robust statistical methodology within the econophysics literature [7]. To this end, we present an econometric investigation of the log-periodic model, that looks closely at residuals properties and deals with the potential difficulties of such models regarding adequate statistical inference. To address these issues, we incorporate both an autoregressive dynamics and a conditional heteroskedasticity structure in the error term of the original model. More specifically, we adopt an AR(1)-GARCH(1,1) structure to explain residual variation across time, these being widely used in the description of the market series in non-crash periods. This means that the extended model, named log-periodic-AR(1)-GARCH(1,1), aggregates some latent dynamical features and mechanisms of the normal phase of the market onto the critical long-range dynamics of price fluctuations encompassed by the original log-periodic model.

The paper is organized as follows. Section 2 focuses on the essentials of the original log-periodic model. In section 3, we present the log-periodic-AR(1)-GARCH(1,1) model and discuss some questions regarding statistical inference in such models. In section 4, the results of model fitting - parameters confidence intervals and residual analysis - are presented for the S&P500 and the NASDAQ (USA). Section 5 concludes the paper and makes some suggestions for future research.

## 2. Econometric Investigation of the Log-periodic Model

We consider the log-periodic model with the following functional form for the time evolution of index prices p(t) prior to crashes [2]:

$$p(t) = A + B(t_c - t)^\beta + C(t_c - t)^\beta \cos[w \log(t_c - t) + \phi] + u_t \qquad (1)$$

where $u_t \sim W(0, \sigma_u^2)$. The deterministic component describes growing oscillations whose period shrinks as the time approaches the critical time $t_c$, identified as the point where the oscillations would accumulate. In Eq. (1), β quantifies the primary power law acceleration of prices and ω measures the frequency of oscillations of the correction term, in logarithmic time units.

In this section, we cover some of the potential econometric problems that one may find when fitting model (1) to index price series. The issues we are interested at are: parameter identifiability, spurious regression, estimation of nonlinear trend models and autocorrelation and heteroskedasticity in the error term. It should be noticed that some of these problems have already been tackled elsewhere (see Ref. [5]).

### 2.1. General premises

Previous empirical studies as well as theoretical fundamentals led to a set of premises on the log-periodic parameters $A, B, C, t_c, \beta, w, \phi$ in (1). The common assumed premises are: $A \sim p(t_c) > 0$ and $B < 0$ for a growing bubble ending in a crash (for β > 0), while $C \neq 0$ guarantees the significance of the log-periodic oscillations. As a major theoretical premise [3], one has $0 < \beta < 1$ for the parameter governing the bubble growth. Empirically, one would request that the log-frequency parameter satisfy 5<ω<15, meaning that there must be some oscillations embedded in the fit to give weight to the model, and otherwise, one should avoid to fit the noise. No previous work mentioned any restriction over $\phi$, but here, we consider $0 < \phi < 2\pi$. With these listed premises, we automatically rule out some obvious non-identifiability [3] problems, while maintaining original interpretations from the log-period framework.

---

[2] For a more general formulation, see Ref. [8].
[3] Following Hamilton [9], we say that a statistical model for a data set **Y** = (Y₁,..., Yₙ)' is *globally identified* for a given vector of parameters ψ iff there is at least one realization **y** of **Y** with

## 2.2. Methodology for parameter estimation

Denote the vector of parameters by $\Phi = [A, B, C, t_c, \beta, w, \phi]$. Under normality assumption for $u_t$, the maximum likelihood estimator $\hat{\Phi}$ is obtained through the minimization of $S(\Phi) = \sum_{t=1}^{n} \hat{u}_t^2(\Phi)$, where $\hat{u}_t(\Phi) = p_t - \hat{p}_t(\Phi)$, with $p_t$, the observed price series and $\hat{p}_t(\Phi)$, the estimated price series according to the specification (1). Hence, the estimates $\hat{\Phi}$ for the log-periodic specification are estimates of ordinary least squares (OLS) for $\Phi$ [9].

Each parameter of the log-periodic model, generally denoted by $\Psi$, defined in a restricted interval denoted by [a, b], is re-parameterized according to the monotonic transformation:

$$\Psi = b \frac{\exp(\varphi)}{1+\exp(\varphi)} + a\left(1 - \frac{\exp(\varphi)}{1+\exp(\varphi)}\right), \qquad \varphi \in (-\infty, +\infty). \tag{2}$$

The transformation (2) turns the original estimation problem over a restricted space of solutions into an unrestricted problem, which is expected to improve the optimization procedure.

The time index $t$ is converted in units of one year. We consider one financial year equal to 252 trading days; thus 1 day = 0.003968 of the year. The time origin is associated with the data of the first observation considered in the series analysis.

Due to the non-linearity of the model, the cost function $S(\Phi)$ is not a strictly convex function, but exhibits a non-trivial landscape, typically with many degenerate or quasi-degenerate local minima. Therefore, the gradient descent method will depend on the starting point fed to the routine. In order to overcome this problem, we apply two complementary optimization algorithms: the Generalized Simulated Annealing (GSA) (see [10]) and the Broyden-Fletcher-Goldfarb-Shanno (BFGS) algorithm (see [11]). The GSA algorithm performs a non-local stochastic search in $\hat{\Phi}$-space while the BFGS algorithm performs, through the descent gradient method, a deterministic fine tuning search. In this work, we evoke the former optimizer to get a "good" initial guess that shall be used within the latter. The GSA algorithm was implemented by means of the C

---

positive probability such that L($\psi$,**y**)≠L($\psi^*$,**y**), where L is the model likelihood function and $\psi^* \in \Theta - \{\psi\}$.

programming language while the BFGS algorithm was implemented using Matlab 7 routines[4].

### 2.3. Stationarity of the residual series

The estimation of statistical models for non-stationary series like the index price movements is potentially problematic, due to the possibility of obtaining spurious regression. In such case, the residual is non-stationary and the parameter estimates might lack statistical meaning. In order to investigate the stationarity of the residual series of the log-periodic specification (1), we perform two unit root (UR) tests: Phillips-Perron (PP) and the Augmented Dickey-Fuller (ADF) [9].

### 2.4. Autocorrelation and heteroskedasticity of the residuals

In a regression framework, there are well-known OLS formulas for parameter estimates and their corresponding standard errors, which are only adequate if the error term in the regression is both homoskedastic and serially uncorrelated. These properties ought to be checked in $\hat{u}_t$, the model residuals. In order to investigate the presence of structures in the residual series, we looked at the residuals autocorrelation function (ACF) and the Ljung-Box test [9]. The presence of linear dependence in the residual series point to the need of incorporating an autoregressive structure AR(1) [9] in the original model.

We must also investigate the presence of non-linear dependence in the residual series by analyzing the ACF for the squared residuals. In order to capture this type of structure, if there is any, a conditional heteroskedastic shock is also added to (1). The parsimonious principle suggests the adoption of a GARCH(1,1) process [9], which has also been widely used in the description of the conditional heteroskedasticity of many financial series.

## 3. The Log-periodic-AR(1)-GARCH(1,1) Model

According to section 2, the empirical findings for the residual of the log-periodic specification applied to financial index series led us to propose a new model to describe the temporal behavior of index prices in the bubble phase, the log-periodic-AR(1)-GARCH(1,1) model:

$$p_t = g(\Phi,t) + u_t. \tag{3a}$$

---

[4] See www.mathworks.com

The deterministic component is given by the log-periodic specification (1) and satisfies the same identifiability conditions of subsection 2.1:

$$g(\Phi,t) = A + B(t_c - t)^\beta + C(t_c - t)^\beta \cos[w\log(t_c - t) + \phi]. \tag{3b}$$

The stochastic component evolves according to an AR(1) process described as

$$u_t = \rho u_{t-1} + \eta_t, \tag{4a}$$

with the error term given by

$$\eta_t = \sigma_t \varepsilon_t, \tag{4b}$$

where $\varepsilon_t$ is a standard random noise term satisfying $E[\varepsilon_t] = 0$ and $E[\varepsilon_t^2] = 1$. The conditional variance $\sigma_t^2$ evolves according to a GARCH(1,1) process:

$$\sigma_t^2 = \alpha_0 + \alpha_1 \eta_{t-1}^2 + \alpha_2 \sigma_{t-1}^2, \tag{4c}$$

where $\alpha_0 > 0, \alpha_1 \geq 0, \alpha_2 \geq 0, \alpha_1 + \alpha_2 < 1$.

### 3.1. Methodology of parameter estimation

Denote the vector parameter of model (3) by $\Theta = [A, B, C, t_c, \beta, w, \phi, \rho, \alpha_o, \alpha_1, \alpha_2]$. The previous difficulty for a suitable estimation of the log-periodic model has grown still further with the incorporation of the AR(1) and GARCH(1,1) structures. The choice of good initial conditions for the optimization routine turns out to be an essential step of the estimation procedure. We propose the following two-stage procedure:

1. the initial guess for the set of parameters $\Phi = [A, B, C, t_c, \beta, w, \phi]$ of the log-periodic component $g(\Phi,t)$ is that from the estimation of the original log-periodic model (1).

2. the initial condition for the set of parameters $(\rho, \alpha_o, \alpha_1, \alpha_2)$ of the stochastic component $u_t$ is drawn back from the residual $\hat{u}_t$ of the original log-periodic model (1), by considering the simultaneous estimates of the AR(1) and GARCH(1,1) structures.

The above procedure's initial guess is used with the conditional maximum likelihood (ML) method. For an n-length series with standard noise $\varepsilon_t \sim N(0,1)$, the general expression for the log-likelihood is [9]:

$$\ln L(\Theta) = -\frac{1}{2}(n-1)\ln(2\pi) - \frac{1}{2}\sum_{t=2}^{n} \ln \sigma_t^2 - \frac{1}{2}\sum_{t=2}^{n} \frac{\eta_t^2}{\sigma_t^2}, \tag{5}$$

where $\eta_t^2$ are $\sigma_t^2$ are obtained under Eqs. (3) and (4).

If the log-periodic AR(1)-GARCH(1,1) model is appropriate for a given financial series, then, when looking at the ACF for both the standardized residual $\hat{\varepsilon}_t = \eta_t^2 / \sigma_t^2$ and its square, one should see no sign of significant autocorrelation left. This being the case, one may conclude that the added AR(1)-GARCH(1,1) structure is able to capture the dependence found in the residual of the original log-periodic specification for the analyzed financial index series.

### 3.2. Residual Diagnostics

Once a log-periodic-AR(1)-GARCH(1,1) model is fitted, one proceeds with usual residual diagnostic tools in order to examine model adequacy. The stationarity of the residuals are investigated using the same set of tests as in section 2.3. The hypothesis of error normality is investigated using the Jarque-Bera (JB) test [9] while the BDS statistics [9] is used to test the null hypothesis of an independent and identically distributed (i.i.d.) residual series. In our applications, due to the small sample sizes, we make use of the bootstrap technique (with 5000 replications) to obtain the correct estimation for the BDS distribution.

### 3.3. Inference in models with a non-linear trend

We then face the problem of possible inferences in a model with non-linear deterministic trend, as it is the case of the specification described by the generating process in Eq. (3). Previous theoretical results [9] for processes with a deterministic linear trend and white noise shock show that the usual $t$ and $F$-statistics have the same asymptotic distributions as those for stationary processes, but with different convergence rates. Unfortunately, similar results for processes with deterministic non-linear trends, as that exhibited by the log-periodic model, are not standard as in the linear case.

In our applications, covariance matrices for the log-periodic-AR(1)-GARCH(1,1) parameter estimators are obtained by inverting the information matrix evaluated for this model. As there are no theoretical grounds fully supporting such variance estimates, some care must be exercised in looking at the inferential results shown in the next section.

# 4. RESULTS

We apply the methodology presented in sections 2 and 3 to investigate the U.S. market, embodying distinct crashes. In Table 1, we show the analyzed index series plus the periods considered for the spanning bubble, which cover an average of two years prior to crashes.

Table 1: Pre-crash analyzed periods of the related index.

| Stock Market | Period |
| --- | --- |
| S&P500 | 01/07/85 to 25/08/87 |
| NASDAQ | 02/01/97 to 10/03/00 |

First, we estimate the log-periodic specification using OLS, according to section 2.2. We investigate the stationarity of the residuals in order to rule out any possible spurious regression. In table 2, we present *p*-values for PP and ADF unit root tests. The results provide strong evidence (even at 1% level) against the unit root null hypothesis of non-stationarity.

Table 2: *P*-values for PP and ADF unit root tests applied to the residuals of the original log-periodic specification.

| | Unit Root tests | | | |
| --- | --- | --- | --- | --- |
| | Phillips-Perron | | ADF | |
| Index | without intercept | with intercept | without intercept | with intercept |
| S&P500 – 1987 crash | 0.0000 | 0.0000 | 0.0000 | 0.0000 |
| NASDAQ – 2000 crash | 0.0002 | 0.0042 | 0.0002 | 0.0041 |

Next, we estimate the log-periodic-AR(1)-GARCH(1,1) specification by means of ML estimations, according to section 3.1, and investigate the stationarity of the residuals. The unit root tests furnish similar results (*p*-values << 0.001), leading us pretty confident that such models have not produced spurious regressions.

## 4.1. S&P500 - 1987 crash

Table 3 presents parameter estimates for the log-periodic specification applied to the S&P500 -1987 pre-crash series. The estimates for β and ω are according to previous findings [2, 12].

Table 3: Log-periodic estimation for S&P500 -1987 pre- crash series.

| SP&500 – 1987 Crash | |
|---|---|
| Parameter | Estimates |
| A | 399.43 |
| B | -153.06 |
| C | -12.09 |
| β | 0.35 |
| ω | 7.28 |
| φ | 1.19 |
| $t_c$ | 2.239 |

Figure 1 displays the time series plot of the S&P500 index prior to the 1987 crash and the optimal curve according to the log-periodic specification given by Eq. (1). In Fig. 2 we show the residuals $\hat{u}_t$ of the estimated model.

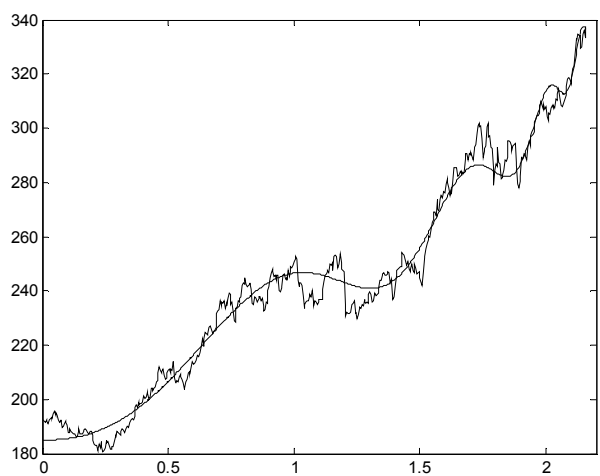

FIG. 1: S&P500 -1987 pre-crash series (ragged) and the log-periodic fit (smooth).

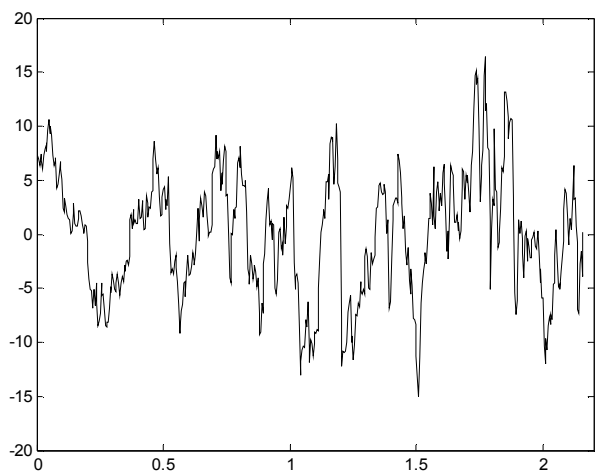

FIG. 2: Residuals of the log-periodic specification
applied to S&P500 - 1985/1987 series.

We now investigate the features of the inner structure of the S&P500 - 1985/1987 residual series exhibited in Fig. 2, according to section 2.4. The ACF for the residual series as well as for the squared residual series reveal the presence of a strong dependence in both residuals. Accordingly, the $Q$-statistic of Ljung-Box for lag $L$, ($Q$-$Stat(L)$), furnish $p$-values $\ll 0.01$ for $L \leq 20$, firmly rejecting the null hypothesis of absence of autocorrelation in both series.

In order to capture the structure observed in the residuals, we perform the estimation of the proposed specification, the log-periodic-AR(1)-GARCH(1,1) model, through the same econometric techniques. Table 4 presents the parameter estimates of Eqs. (3) and (4) applied to the S&P500 -1987 pre-crash series.

Table 4: Log-periodic-AR(1)-GARCH(1,1)
estimation for S&P500 -1987 pre- crash series.

| S&P500 – 1987 crash | |
|---|---|
| Parameter | Estimates |
| A | 385.11 |
| B | -141.15 |
| C | -12.04 |
| β | 0.37 |
| w | 6.97 |
| φ | 1.41 |
| ρ | 0.935 |
| $α_0$ | 0.023 |
| $α_1$ | 0.036 |
| $α_2$ | 0.962 |
| $t_c$ | 2.210 |

Comparing tables 3 and 4, one verifies that the log-periodic parameters are rather robust against the incorporation of the residual structure. On the other hand, the estimated critical time $t_c$ of the new model approaches the actual crash starting time $t_{max}$ (the price drawdown started at $t_{máx}$ =2.159 and reached the lowest value at $t_{min}$ =2.310, in converted time unit – see section 2.2).

Figure 3 shows the temporal series of the standardized residuals $\hat{\varepsilon}_t$ obtained from the estimated model for the S&P500 -1985/1987 series.

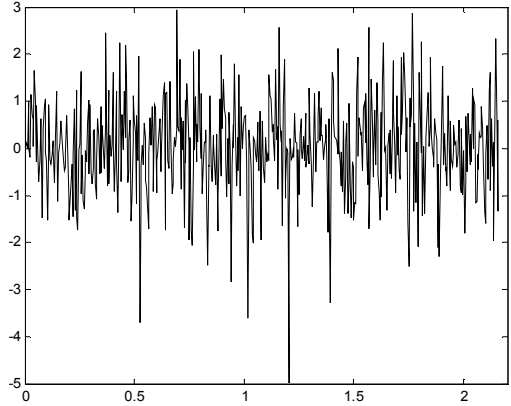

FIG. 3: S&P500 -1985/1987 residuals of the log-periodic-AR(1)-GARCH(1,1) specification.

We proceed with the ACF analysis of the standardized residuals of the log-periodic-AR(1)-GARCH(1,1) specification. From Table 5, the *Q-Stat(20)* for the residuals and the squared residuals has *p*-value 0.586 and 0.836 respectively, providing strong evidence to accept the null hypothesis of absence of autocorrelation of first and second order in the residual series of the extended model.

Carrying on the investigation of the standardized residuals $\hat{\varepsilon}_t$, we perform the descriptive statistics of the residuals and the JB test for examining normality. From Table 5, the residual distribution suitably presents zero mean and unitary variance but the kurtosis evaluation and the JB test indicates a non-Gaussian character of the residual series.

In table 6, we show the results for the BDS statistics. The confidence levels are obtained from the bootstrap technique with 5000 replications, according to section 3.2. The null hypothesis is strongly accepted, supporting that the standardized residuals of the log-periodic-AR(1)-GARCH(1,1) model applied to the S&P500- 1985/ 1987 series are i.i.d..

Table 5: Descriptive statistics, JB test and Q-statistic for the S&P500-1985/1987 standardized residuals of the log-periodic-AR(1)-GARCH(1,1) specification (*p*-values in parenthesis).

|  | S&P500 – 1987 crash |
|---|---|
| Mean | 0.017 |
| Standard deviation | 0.988 |
| Skewness | -0.384 |
| Kurtosis | 4.774 |
| Jarque-Bera test | 84.673 (0.00) |
| Q-Stat(20) – residuals | 18.023 (0.586) |
| Q-Stat(20) – squared residuals | 13.884 (0.836) |

Table 6: *P*-values of the BDS statistic for the S&P500 - 1985/1987 standardized residuals of the log-periodic-AR(1)-GARCH(1,1) specification.

| M \ ε | 0.5σ | 1.0σ | 1.5σ | 2.0σ |
|---|---|---|---|---|
| 2 | 0.92120 | 0.78320 | 0.75120 | 0.85120 |
| 3 | 0.70200 | 0.94880 | 0.97040 | 0.86440 |
| 4 | 0.79320 | 0.96400 | 0.99160 | 0.76360 |
| 5 | 0.57560 | 0.92400 | 0.87200 | 0.47920 |
| 6 | 0.57240 | 0.90920 | 0.98160 | 0.66520 |

The above results indicate that the estimates have statistical meaning, allowing an inference statistical analysis of the coefficients. In table 7, we present the results of this analysis, namely the standard error, the t-statistic and the confidence interval (CI) at the 95% level.

Table 7: Statistical inference analysis of the log-periodic-AR(1)-GARCH(1,1) parameters for S&P500 - 1987 pre-crash series.

| Parameter | Coefficient | Standard error | t-statistic | CI lower | CI upper |
|---|---|---|---|---|---|
| A | 385.11 | 25.62 | 15.03 | 334.90 | 435.32 |
| B | -141.15 | 26.31 | 5.37 | -192.71 | -89.59 |
| C | -12.04 | 1.69 | 7.12 | -15.36 | -8.73 |
| β | 0.37 | 0.075 | 5.00 | 0.23 | 0.52 |
| w | 6.97 | 0.375 | 18.59 | 6.23 | 7.70 |
| φ | 1.41 | 0.224 | 6.31 | 0.97 | 1.85 |
| ρ | 0.935 | 0.016 | 57.708 | 0.903 | 0.967 |
| $α_0$ | 0.023 | 0.02 | 1.031 | - | - |
| $α_1$ | 0.036 | 0.016 | 2.218 | 0.004 | 0.07 |
| $α_2$ | 0.962 | 0.018 | 54.67 | 0.93 | 1.00 |
| $t_c$ | 2.210 | 0.013 | 165.92 | 2.183 | 2.237 |

The t-statistic tests the null hypothesis that the coefficient is zero with asymptotic standard Normal distribution. From table 7, all the tests and inferences points towards confident and consistent coefficients, the only exception being $\alpha_0$.

### 4.2. NASDAQ – 2000 crash

Figure 4 presents the time series plot of the NASDAQ index prior to the 2000 crash and the optimal curve according to the log-periodic specification.

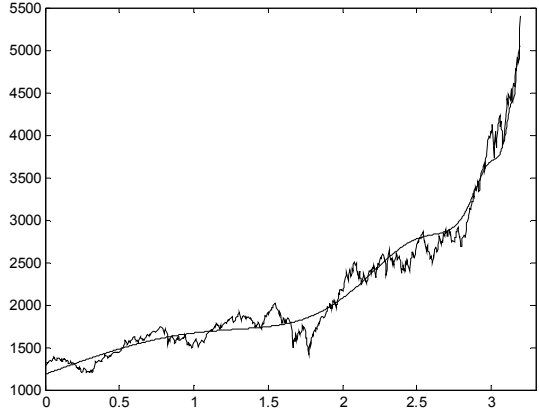

FIG. 4: Nasdaq-2000 pre-crash series (ragged) and the log-periodic fit (smooth).

The estimates according to the extended models are shown in Table 8. The estimated $t_c$ are within a few days from the actual crash starting date (the price drawdown started at $t_{máx}$=3.194 and reached the lowest value at $t_{min}$=3.397, in converted time unit).

Table 8: Log-periodic-AR(1)-GARCH(1,1) estimation for NASDAQ – 2000 pre-crash series.

| Parameter | Log-per-AR(1)-GARCH(1,1) |
|---|---|
| A | 5619.2 |
| B | -3247.5 |
| C | 151.58 |
| $\beta$ | 0.27 |
| $\omega$ | 5.65 |
| $\phi$ | 1.56 |
| $\rho$ | 0.972 |
| $\alpha_0$ | 10.722 |
| $\alpha_1$ | 0.198 |
| $\alpha_2$ | 0.817 |
| $t_c$ | 3.200 |

In Table 9, we investigate the standardized residuals $\hat{\varepsilon}_t$. Excepting for the normality assumptions, the results show an adequate fit and absence of first and second order correlations in the residuals. In addition, the BDS test results shown in table 10 strongly support the i.i.d. hypothesis.

Table 9: Descriptive statistics, JB test and Q-statistic for the NASDAQ - 1997/2000 standardized residuals of the log-periodic-AR(1)-GARCH(1,1) specification (*p*-values in parenthesis).

|  | NASDAQ – 2000 crash |
|---|---|
| Mean | -0.023 |
| Standard deviation | 0.997 |
| Skewness | -0.539 |
| Kurtosis | 3.993 |
| Jarque-Bera test | 72.107 (0.000) |
| Q-Stat(20) – residuals | 20.309 (0.439) |
| Q-Stat(20) – squared residuals | 14.062 (0.827) |

Table 10: *P*-values of the BDS statistic for the NASDAQ- 1997/2000 standardized residuals of the log-periodic-AR(1)-GARCH(1,1) specification.

| M \ ε | 0.5σ | 1.0σ | 1.5σ | 2.0σ |
|---|---|---|---|---|
| 2 | 0.53800 | 0.44920 | 0.15760 | 0.21480 |
| 3 | 0.93880 | 0.83520 | 0.87600 | 0.87560 |
| 4 | 0.96280 | 0.90760 | 0.98160 | 0.67280 |
| 5 | 0.93040 | 0.94280 | 0.98800 | 0.57400 |
| 6 | 0.87520 | 0.89120 | 0.91640 | 0.51000 |

We now present the results of the statistical inference for the coefficients of the log-periodic-AR(1)-GARCH(1,1) model applied to the NASDAQ- 1997/2000 series in table 11. All tests and inferences lead to confident and consistent coefficients, with the exception of $\alpha_0$.

Table 11: Statistical inference analysis of the log-periodic-AR(1)-GARCH(1,1) parameters for Nasdaq-2000 pre-crash series.

| Parameter | Coefficient | Standard error | t-statistic | CI lower | CI upper |
|---|---|---|---|---|---|
| A | 5619.2 | 799.52 | 7.028 | 4052.13 | 7186.27 |
| B | -3247.5 | 824.89 | 3.937 | -4864.28 | -1630.72 |
| C | 151.58 | 41.42 | 3.659 | 70.39 | 232.77 |
| β | 0.27 | 0.079 | 3.487 | 0.12 | 0.43 |
| ω | 5.65 | 0.215 | 26.317 | 5.23 | 6.07 |
| φ | 1.56 | 0.225 | 6.951 | 1.12 | 2.00 |
| ρ | 0.972 | 0.009 | 113.655 | 0.96 | 0.99 |
| $\alpha_0$ | 10.722 | 5.74 | 1.869 | - | - |
| $\alpha_1$ | 0.180 | 0.045 | 4.390 | 0.09 | 0.27 |
| $\alpha_2$ | 0.817 | 0.039 | 21.136 | 0.74 | 0.89 |
| $t_c$ | 3.200 | 0.0023 | 1411.589 | 3.195 | 3.205 |

### 4.3. Estimated crash time: comparison with previous results

It is worth to mention that, as the estimated models may depend on the optimization procedure, the comparison among the point estimated parameters presented in the literature should be taken with caution. Particularly, the fits have been performed in different time units, which prevents the direct comparison of the critical time of the crash $t_c$. Moreover, confidence interval ought to be used in the comparison.

With this aim, we present in Table 12 the calendar date of the estimated $t_c$ for the log-periodic-AR(1)-GARCH(1,1) model (confidence interval) and compare it with previous estimates for the original log-periodic model. The results for the analyzed series suggest that the introduction of the residual structure improves the estimated crash time $t_c$.

Table 12: Comparison of the estimated crash date $t_c$.
(the price drawdown starts at $t_{máx}$ and reaches the lowest value at $t_{min}$).

|  | S&P 500 - 1987 crash | NASDAQ - 2000 crash |
|---|---|---|
| Date $t_{máx}$ | 25-Aug-87 | 10-Mar-00 |
| Date $t_{min}$ | 19-Oct-87 | 23-May-00 |
| Date $t_c$ (this work) | 4-Set-87 to 18-Set-87 | 12-Mar-00 to 16-Mar-00 |
| Date $t_c$ (previous results) | 01-Oct-87 [2] 27-Set-87 [12] | 31-Mar-00 [13] |

## 5. Conclusion

Although several econometric models have been successful in describing the stylized facts observed in financial time series during normal regimes of the market, a model that suitably describes the temporal behavior of the speculative bubbles before crashes is still lacking. The present work aimed to contribute for the achievement of proper models for the pre-crash regime of prices, satisfying statistical inference criteria.

To this end, we first perform a detailed statistical analysis of a class of log-periodic models that has recently been proposed to describe the temporal behavior of prices before crashes. We have analyzed the most simple specification, given in Eq.(1), applying it to the financial index series of the U.S. market. The outline of our econometric investigation is that, although the estimated residuals of the log-periodic specification are stationary, they exhibit strong autocorrelation and heteroskedasticity.

For this kind of inner structure in the residuals, the present "state of art" of the econometric analysis does not provide consistent asymptotic estimators for the variance of the regression estimators. Moreover, due to the non-linear character of the model, this analysis turns out to be even subtle. Considering the above arguments, one leads to the conclusion that the original log-periodic model fails in providing confident parameter estimates, even though, as illustrated here for the analyzed series, the optimum fits of the model suitably describes the mean function of the index price series.

The non-trivial inner structure of the residuals reveals that the log-periodic model did not exhaust the full complexity of the bubble regime. On the other hand, the stationarity of the residuals encourages the investigation of possible extensions of the model that would lead to reliable statistical inferences.

The econometric analysis of the proposed log-periodic-AR(1)-GARCH(1,1) model shows that, for the examined financial index series, the residuals are independent and identically distributed. This means that the new model was able to capture the inner structure found in the residual of the original log-periodic specification.

The attainment of independent and stationary residuals is a necessary condition towards parameter estimates with statistical meaning. Nevertheless, rigorous expressions for the asymptotic test statistics in the case of models with non-linear trend are still lacking. In this work, we consider that previous theoretical analysis for models with linear trend may be extended for the log-periodic-AR(1)-GARCH(1,1) model. Based on this assumption, our statistical inference analysis leads in general to consistent and confident parameters, validating our approach.

The results for the analyzed series suggest that the log-periodic parameters are rather robust against the introduction of the residual structure, while the estimated critical time $t_c$ is improved as compared with previous results for the original model. This forecasting advantage should be tested for incoming crashes.

There are some possible extensions for the analysis carried out in this paper, as for instance, the study of the true asymptotic distribution for the ML estimators in the case of models with non-linear trend, as our log-periodic AR(1)-GARCH(1,1) model, by Monte Carlo simulation. Another issue that arises is the one concerning the adoption of heavier tail distributions for the error term of the model, such as t-Student.

# Acknowledgments

This work has been partially supported by the Brazilian agencies FAPERJ , CAPES and CNPq.